\documentclass[useAMS,usenatbib,fleqn]{mn2e}

\usepackage{txfonts}
\usepackage[pdftex]{graphicx}
\usepackage{graphicx}
\usepackage{natbib}
\usepackage{subfigure}

\usepackage{ulem}
\usepackage{xcolor}
\usepackage{pstricks}
\usepackage{verbatim}
\bibpunct{(}{)}{;}{a}{}{,}

\usepackage{times}

\title{Brightest group galaxies and the large--scale environment}

\author[Luparello et al.]{%
    H.~E.~Luparello$^{1}$,
    M.~Lares$^{1}$,
    D.~Paz$^{1}$, 
    C.Y.~Yaryura$^{1}$, 
    D.~G.~Lambas$^{1}$ and
    N.~Padilla$^{2,3}$\\
$^{1}$Instituto de Astronom\'{\i}a Te\'{o}rica y Experimental
  (CONICET-UNC). Observatorio Astron\'{o}mico de C\'{o}rdoba, 
   C\'{o}rdoba, Argentina\\
$^{2}$Departamento de Astronom\'{\i}a y Astrof\'{\i}sica, Pontificia
   Universidad Cat\'olica de Chile, Santiago, Chile\\
$^{3}$Centro de Astro-Ingenier\'{i}a, Pontificia Universidad Cat\'olica 
   de Chile, Santiago, Chile}

\date{Released 2014 Xxxxx XX}
\pagerange{\pageref{firstpage}--\pageref{lastpage}} \pubyear{2014}

\def\LaTeX{L\kern-.36em\raise.3ex\hbox{a}\kern-.15em
    T\kern-.1667em\lower.7ex\hbox{E}\kern-.125emX}

\begin{document}
\label{firstpage}
\maketitle

\begin{abstract}
We study the dependence of the properties of group
galaxies on the surrounding large--scale environment, using SDSS-DR7
data.
Galaxies are ranked according to their luminosity within each group and
classified morphologically by the S\'ersic index.
We have considered samples of the host groups in superstructures of
galaxies, and elsewhere.
We find a significant dependence of the properties of late--type
brightest group galaxies on the large--scale environment: they show
statistically significant higher luminosities and stellar masses,
redder \textit{u-r} colours, lower star formation activity and longer
star--formation time--scale when embedded in superstructures. 
By contrast, the properties of the early--type brightest group galaxies
are remarkably similar regardless of the group global environment. 
The other group member galaxies exhibit only the local influence of the
group they inhabit.
Our analysis comprises tests against the dependence on the host group
luminosity and we argue that group brightest member properties are not
only determined by the host halo, but also by the large--scale
structure which can influence the accretion process onto their
late--type brightest galaxies.
%
\end{abstract}

\section{Introduction} \label{S_intro}
%
The observed distribution of galaxies at large scales shows a complex network
of filaments and voids \citep[e.g.][]{Joeveer:1978, Zeldovich:1982, Einasto:1996, Colless:2001,
Jaaniste:2004, Einasto:2006, Abazajian:2009}.
After extended galaxy surveys \citep{York:2000, Colless:2001} were
completed, this picture was widely studied and confirmed, and it is
also supported by the analysis of several numerical simulations
consistent with the standard cosmological model \citep{Frisch:1995,
Bond:1996, Seth:2003, Shandarin:2004}.
The hierarchical structure formation model states that the
distribution of matter evolves from small fluctuations in the early
universe to form the complex observed large--scale configuration
\citep[see e.g.][]{PS:1974, Bardeen:1986, White:1978}.
During this evolutionary gravitational process, galaxies initially
flow from underdense to higher density regions
\citep[e.g.][]{Gunn:1972}.
Thus, the formation of voids and superstructures can be considered as
complementary processes retaining useful information to infer the
cosmological parameters.
The higher density regions, usually named superclusters, are
preferentially located at the intersections of walls and filaments.
The galaxy superclusters are the largest systems of the
Universe, hosting a variety of galaxy systems such as groups and clusters of galaxies.

It is widely accepted that galaxy properties are strongly influenced
by the local environment in which they reside
\citep{Blanton:2003b,Balogh:2004,Kauffmann:2004,Wilman:2005,Baldry:2006,
Hou:2009,Peng:2010,McGee:2011,Patel:2011,Sobral:2011,Muzzin:2012,Hou:2013}.
However, recent studies have analysed the influence of large--scale
structure over the systems that they host. 
It is well known that more luminous groups and clusters are located in
higher density environments \citep{Einasto:2003b, Einasto:2005,
Croft:2012, Yaryura:2012, Luparello:2013}. 
Also, galaxy properties in these systems such as the star
formation rate and colors depend on the large--scale structure
\citep{Binggeli:1982, Donoso:2006, Einasto:2007c, Crain:2009,
White:2010}.
\citet{Lietzen:2012} analyse galaxies in groups inhabiting different
large--scale environments in the eighth data release (DR8) of the Sloan
Digital Sky Survey \citep[SDSS,][]{York:2000}.
They conclude that in high--density environments the amount of passive
elliptical galaxies rises while the star--forming spirals decrease.
Thus, considering equally rich groups, they contain a large fraction of elliptical galaxies
when located in supercluster environments.

Within this framework, \citet{Einasto:2014} confirm that local and
global environments have a significant impact over the formation an
evolution of galaxies.
They also find that even supercluster morphologies influence galaxy
properties.
Using Minkowski functionals they characterize supercluster
morphologies spliting them into spider and filament types.
Filament--type superclusters contain a larger fraction of red, early
and low star--forming galaxies than spider--type. 
Besides, red and star--forming galaxies hosted in spider--type
superclusters present higher stellar masses than in filament--type. 
Moreover, equally rich groups located in filament--type superclusters
contain a large fraction of red, early--type galaxies, with large
stellar masses and low star--formation rates. 

\citet{Luparello:2013} study differences between galaxy group
properties according to the large--scale environment over the seventh
data release (DR7) of the SDSS.
They assert that the total galaxy density profile around groups is
independent of the global environment.
However, groups in superstructures have larger stellar mass content,
higher velocity dispersions and older stellar populations than groups
elsewhere.
As they compare equal global luminosity groups, a proxy of the total
mass, they conclude that groups in superstructures formed earlier than
groups located in lower density regions.
In the same line, \citet{Chon:2014} provide evidence that the special
environment of superclusters, as characterized by the X--ray luminosity
function, originates from a top--heavy mass function.

The degree to which the large--scale structure of the universe determines
 properties of galaxies, irrespective of the local environment, is still
a promising field to understand many aspects of galaxy formation
and evolution.
This subject has been recently addessed by a number of works, either
on galaxy survey data \citep{tempel_galaxy_2011,
einasto_superclustersIII_2007} or focusing on particular known
superclusters, like Ursa Major \citep{Krause:2013}, Coma
\citep{Cybulski:2014}, Shapley \citep{Merluzzi:2014} Sculptor and the
Sloan Great Wall \citep{einasto_toward_2008}.
For example, \citet{Krause:2013} analyse the distribution of galaxy
groups in the Ursa Major Supercluster, finding that relaxed systems
around high density peaks may have formed and evolved earlier than
nonrelaxed systems, which are growing slower on the peripheries of
lower density peaks.
In another noteworthy study, \citet{Cybulski:2014} explore the
star--forming activity of galaxies located in different regions of the
Coma Supercluster, concluding that the fraction of star--forming
galaxies progressively decreases from voids to filaments.         
According to the authors, the fraction of blue galaxies declines as
the environmental density rises.  

\citet{einasto_toward_2008} present a study of the Sculptor
supercluster and the Sloan Great Wall, two of the richest
superclusters in the 2dF Galaxy Redshift Survey.
They find differences in the galaxy content according to the local
density within the supercluster, with a smaller fraction of early--type
galaxies towards the outskirts.
A related study carried out on the 2dF redshift survey
\citep{einasto_superclustersIII_2007} shows evidence of a larger
fraction of early--type galaxies in superclusters with a larger global
density.

\citet{tempel_galaxy_2011} state
that the luminosity function of elliptical galaxies strongly depends
on the environment, while the luminosity distributions of late--type
galaxies are similar.
More recently, The Shapley Supercluster Survey \citep[ShaSS,][]{Merluzzi:2014}
which covers a region of $260Mpc^3$, including the Shapley supercluster core
and mapping nine Abell clusters and two poor clusters,
searches for the role of cluster--scale mass
assembly on galaxy evolution and possible connections between
the properties of the cosmic structures.

In this paper we analyse different properties of galaxies in groups in order
to assess their dependence on the surrounding large--scale structure.
We have studied the correlation of galaxy properties, namely: luminosities, 
colours, star--formation rates, concentration index and time--scales,
and the large--scale environment constraing
the local environment through the selection of galaxy groups.
This paper is organized as follows. 
In Section \ref{S_data} we describe the data samples of galaxies, galaxy groups, and superstructures.
We also describe in this section the use of S\'ersic index to discriminate galaxy morphologies
and the ranking of galaxy luminosities within each group.
In Section \ref{S_properties} we analyse the properties of the brightest group galaxies and in Section \ref{S_BGGtype} we 
address their dependence on large--scale environment.
We summarize the results and state the main conclusions in Section \ref{S_conclusus}. 
Throughout this paper, we adopt a concordance cosmological
model($\Omega_{\Lambda}=0.75$, $\Omega_{matter}=0.25$).

\section{Data and Samples} \label{S_data}
%
\subsection{SDSS--DR7 Galaxy Catalogue} \label{SSDS-DR7} 
In the present work we use the Seventh Data Release \cite[DR7,][]{Abazajian:2009} of the 
Sloan Digital Sky Survey \citep{York:2000}, which is publicly available 
\footnote{http://www.sdss.org/dr7}.
The footprint area comprised by the spectroscopic galaxy catalogue is 9380 sq.deg and 
its limiting apparent magnitude in the r--band is 17.77 \citep{Strauss:2002}.
We use a more conservative limit of $17.5$ to ensure completeness in our samples,
and in order to avoid saturation effects in the photometric pipeline, we
consider galaxies fainter than \mbox{r = 14.5} 
\footnote{http://www.sdss.org/dr7/products/general/target\_quality.html}.
We also use the value added galaxy catalog from MPA-JHU DR7 \citep{Kauffmann:2003}, 
which provides additional information about star--formation rates and stellar masses.
The star--formation rates (SFRs) are computed following the procedure 
described by \citet{Brinchmann:2004}, while the stellar masses are obtained 
as explained in \citet{Kauffmann:2003} and \citet{Salim:2007}.

The one-component S\'ersic fits to galaxy radial profiles
can be used to roughly estimate the morphological galaxy clasification
\citep{Blanton:2003b}.  We use S\'ersic indices from 
the NYU Value-Added Galaxy Catalog \cite[NYU-VAGC,][]{Blanton:2005}
to separate two samples of galaxies with different morphological
properties.  
Details about the chosen threshold values of the S\'ersic
indices and the properties of the subsamples 
are given in subsection \ref{galaxy_samples}.

\subsection{SDSS--DR7 Superstructures} \label{superstructures} 
Galaxies in the universe are arranged in a supercluster--void network.
Superclusters enclose a wide range of galaxy structures ranging from single galaxies to rich clusters.
Thus, these environments may provide important clues to unveil galaxy
formation physics, evolution and large--scale clustering.

There are several supercluster catalogues \cite[e.g.,][]{Einasto:2007,
CostaDuarte:2010, Luparello:2011, Liivamagi:2010}
constructed using the so called density field method.
However, as these systems are going through their virialization
process, there is a certain degree of freedom in defining the applied
density threshold.
According to the \mbox{$\Lambda$CDM} Concordance Cosmological Model,
the present and future dynamics of the universe are dominated by an
accelerated expansion.
By combining the luminosity density field method \citep{Einasto:2007}
with the theoretical criteria for the mass density threshold of
\cite{Dunner:2006}, \citet{Luparello:2011} presented a catalogue of
so called Future Virialized Structures (FVS).
The density field method allows to compute the galaxy luminosity
density field by convolving the spatial distribution of galaxies with
a kernel function weighted to the galaxy luminosity.
We used an Epanechnikov kernel of \mbox{8 $h^{-1}\,Mpc$} to sample the
density field into a redshift--space grid composed by
cubes of \mbox{1 h$^{-1}$ Mpc} side.
We then select the highest luminosity density groups of cells to
isolate the large structures.
However, the exact value of the luminosity density threshold has been
a matter of some debate, since different values give rise to entirely
different samples of superstructures.
In a previous work \citep{Luparello:2011}, we used a criterion motivated by
dynamical considerations to calibrate this value.
The resulting large--scale structures
correspond to overdense regions in the present-day universe that will
become virialized structures in the future.
The catalogue of FVS was compiled using a volume--limited sample of galaxies
from the SDSS--DR7, in the redshift range \mbox{$0.04<z<0.12$}, with a
limiting absolute magnitude of \mbox{$M_r <-20.47$}. 
According to calibrations made using mock
catalogues, the final sample of FVS is 90 per cent complete and has contamination
below 5 per cent.
The volume covered by the catalogue is \mbox{3.17 $\times$ $10^7$
(h$^{-1}$ Mpc)$^3$}, within which 150 superstructures were identified,
composed by a total of 11394 galaxies.
FVS luminosites vary between \mbox{ $10^{12}$ L$_{\odot}$} and
\mbox{$\simeq$ 10$^{14}$ L$_{\odot}$}, and their volumes range between
\mbox{10$^2$ (h$^{-1}$ Mpc)$^3$} and  \mbox{10$^5$ (h$^{-1}$
Mpc)$^3$}.

%
\subsection{SDSS--DR7 Group Catalogue} \label{galaxy_groups} 
%
The clustering properties and the formation and evolution processes of
mass in the scale of galaxy groups are a result of the hierarchical
accretion of mass that also gives rise to luminous galaxies.
For that reason, galaxy groups are key to observationally constrain
those processes.
To characterize the local environment of galaxies we use galaxy groups
identified by \cite{Zapata:2009} in the SDSS galaxy catalogue,
extended to cover the SDSS--DR7.
The identification method is based on a ''Friends of Friends''
algorithm, which is one of the most commonly used percolation
algorithms.
It allows to tie in sets of galaxies, where each galaxy is closer than
a given linking length to at least another galaxy in the group.
The parameters in the algorithm can be adjusted so that the resulting
groups resemble the regions occupied by dark matter haloes. 
Following \citet{Merchan&Zandivarez:2005}, \citet{Zapata:2009}
 used a variable
projected linking length and a fixed radial linking length, in order
to compensate the decrease of the number of galaxies with redshift in
flux--limited samples.
The parameters of the percolation algorithm have been calibrated using
mock catalogues, so that an optimal compromise between completeness
and contamination is achieved.
The varying projected linking length $\sigma=\sigma_{0}\times R$ has
a value of $\sigma_{0}=0.239h^{-1}Mpc$ (where R is the scaling factor as defined
in equation (4) of \citet{Merchan&Zandivarez:2005}), and the fixed radial
linking length is $\Delta v=450kms^{-1}$.
The configuration adopted for the FoF identification of galaxy groups corresponds
to the values calibrated by \citet{Merchan&Zandivarez:2005} to obtain 95 per
cent of completeness and 8 per cent of contamination.
The identificaction is performed on a flux limited galaxy sample.
The catalogue contains 83784 groups with at least 4 members, and is
limited to redshift \mbox{$z < 0.2$}.
%

\subsection{SDSS--DR7 Galaxy and Group samples} \label{galaxy_samples} 
With the aim to study the impact of large--scale environment over
galaxy properties, we analyse the variation of characteristic
parameters of galaxies in groups inside and outside superstructures.
First we select galaxy groups within the volume of the FVS catalogue
with multiplicities between 5 and 15 members.
We use a lower multiplicity limit of five members to diminish
contamination and an upper limit of fifteen members taking into account Figure 2 of
  \citet{Luparello:2013}, who find that  for $n>15$ the relation between multiplicity and
  group luminosity differs significantly between groups whithin FVS and
  elsewhere.

In addition, we compute total group luminosities using the volume--limited sample of
galaxies in groups within the limiting redshift.
The limiting absolute magnitude in the r--band corresponding to the limiting redshift
$z=0.12$ is $M_{rlim}=-20.47$.
The total group luminosities (which can be considered as a proxy of group mass) were obtained by adding the r--band luminosities of the
member galaxies brighter than $M_{rlim}$.

FVS are globally overdense systems, but they also have complex
morphologies and their different regions present a wide range of density
levels.    
To ensure that we are analizing realiable samples, we select groups
located in the densest cores of superstructures and not around their
peripheral zones.
In order to accomplish this, we estimate the mean luminosity density
on a 13 $h^{-1}Mpc$ side cube (which corresponds aproximately to the volume of
a sphere of radius 8 $h^{-1}Mpc$) centered on each group.  
We keep groups with mean luminosity density equal or greater
than 5.3 times the mean overall luminosity density estimated on the
SDSS--DR7, $\bar{\rho}=1.73\times 10^8 L_{\odot}$.
This threshold was chosen according to the calibration previously
presented by \citet{Luparello:2011}.  
In order to have a suitable number of
groups per luminosity bin, we also restrict group luminosities to the range
$10^{10.9}$--$10^{11.35}L_{\odot}$.
Under these conditions, we obtain a sample of 123 groups inhabiting
the densest regions of FVS.
For an adequate comparison, we also select groups not belonging to
FVS. 
By appliying the same restrictions over group members and
luminosities, we obtain a sample of 372 groups outside FVS.
Regarding the galaxies that conform the group samples, we remove those
with magnitude uncertanties greater than $0.05$ in the r--band.
The final samples comprise
861 galaxies in groups located in FVS
and 2620 galaxies in groups out of FVS, in the redshift range $0.04 \le z \le 0.12$.
\subsection{Luminosity gap in groups} \label{lgap} 

We define the "luminosity ranking" of galaxies with respect to their host
group, arranging them in descending order of r--band luminosity, then the
brightest group galaxy (BGG) is the first ranked.
The BGG properties and their relation with the environment have been
widely studied.
\citet{DeLucia:2012} use galaxy merger trees from semi-analytic model
simulated catalogues to study the histories of galaxies in groups
according to their environment.
In a hierarchical structure formation driven picture of the assembly
of groups, the authors claimed that there is a history bias that shapes
the properties of central and satellite galaxies and that to some
extent depends on the large--scale structure.
\citet{Shen:2014} establish that BGGs are more luminous than expected
from the ordered statistics, indicating that its distinctive
brightness may be a consequence of physical processes rather than just
an artifact of them being defined as the brightest galaxy in its
group.
They state that this brightening process, produced by the growth of
stellar mass, may be associated to local processes.

When the gap in luminosity between the BGG and the rest of the members
in the group is large, the system can be considered as a dominant
galaxy with satellites.
This type of systems has been studied to assess the role of the
primary galaxy on the evolution of the group.
By analysing the properties of faint satellites associated to isolated
bright galaxies, \citet{Lares:2011} conclude that the satellite
overdensity depends on the luminosity and color of the primary galaxy
and on the luminosity of the satellites.
These systems of satellites around bright isolated galaxies 
are found to be more concentrated and more populated for red, passive galaxies, 
and for larger stellar mass central galaxies
\citep{Agustsson:2010,Guo:2011,Lares:2011,Wang:2012,Wang:2014}.
Given this correlation, early--type brightest group galaxies are expected to
reside in more dynamically relaxed groups compared to those where the
BGG is of late--type.  

\subsection{Group samples and galaxy classification} \label{galaxy_samples} 

As stated above, BGG properties are strongly related to the intrinsic properties
of their host groups.
Taking this into account, we use the BGG characteristics to
distinguish between different host group classes.
Morphological classifications of galaxies have been linked to their
formation and evolution \citep{Pannella:2009}, as well as to their
star formation  and central black hole activity
\citep{Schawinski:2010}.
Also, galaxy morphological types depend on the surrounding environment
\citep{Bamford:2009}.

The radial dependence of surface brightness of galaxies can be
considered as an indicator of their morphology \citep{Trujillo:2001b}.
The one-component S\'ersic fits \citep{Sersic:1963, Sersic:1968} to
galaxy radial profiles can be used to roughly estimate the
morphological galaxy clasification \citep{Blanton:2003b}.
The S\'ersic profile,
\begin {equation} I(r) = I_{0} \exp[-(r/r_0)^{1/n}], \end {equation}
is parameterized by $I_0$, the central surface--brightness, $r_0$, the
scale radius, and $n$, the profile index for each galaxy.
The S\'ersic index $n$ is correlated to the morphological type: $n =
4$ represents the profile of elliptical galaxies $r^{1/4}$
\citep{deVaucouleurs:1948} while $n = 1$ corresponds to the
exponential profile of spiral disk galaxies. 
There are previous studies based on SDSS which consider the S\'ersic
index as a morphological indicator.
Even though most of these studies assume empirical cuttoffs around $n
= 2.5$ to establish morphological distinction, this value depends on
the specific aim of the analysis.
For instance, \citet{Blanton:2003b} divide the SDSS galaxies into two
groups according to this index.
They distinguish an exponential group with $n < 1.5$ and a
concentrated group with $n > 3$, analysing the relationships between
galaxy parameters.
Furthermore, \citet{Shen:2003} choose $n = 2.5$ which is the average
between exponential and the Vaucouleurs profiles, while
\citet{Hogg:2003} apply $n = 2$.

\begin{figure} 
   \centering
   \includegraphics[width=0.45\textwidth]{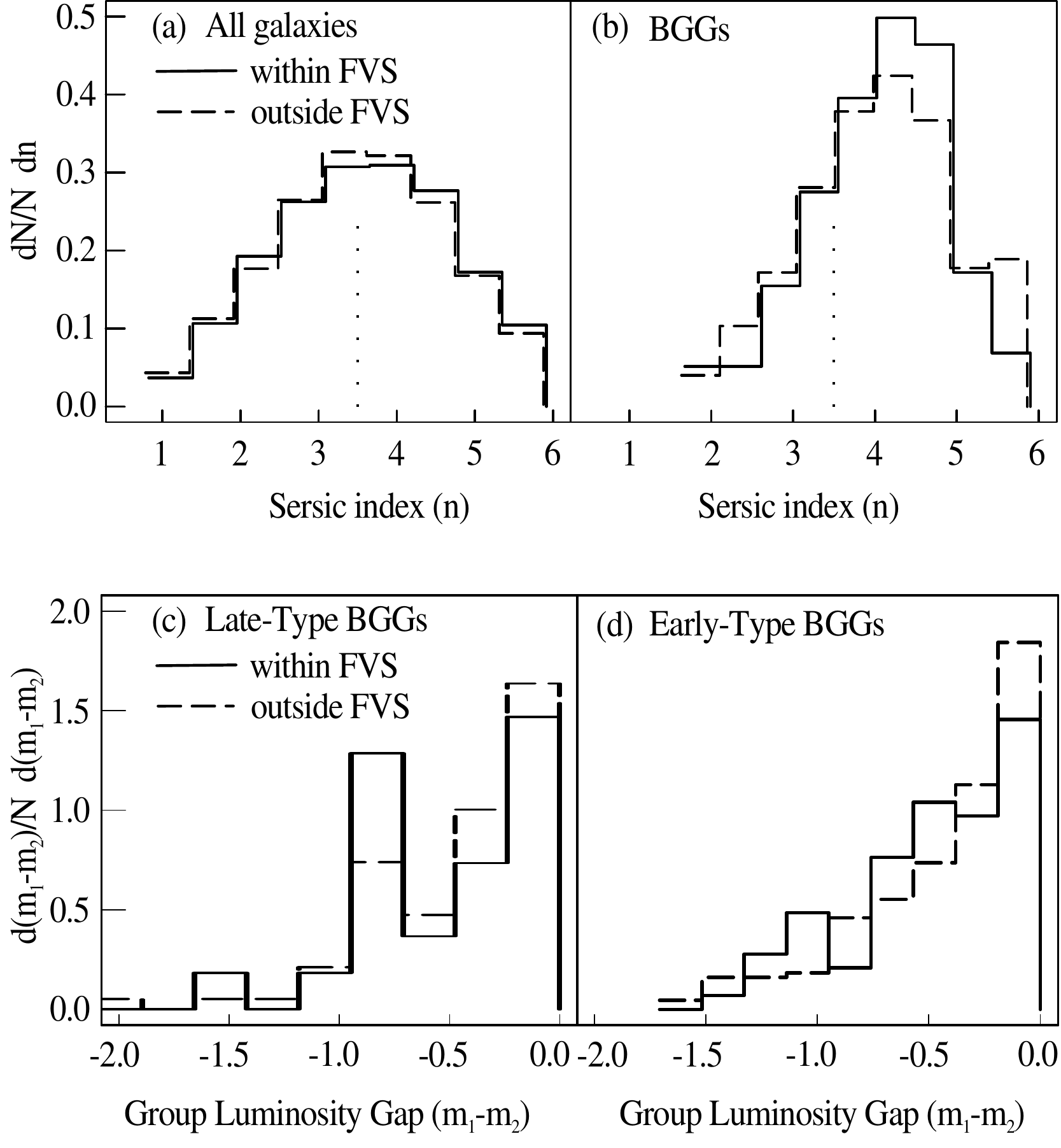}
   \caption{(a) S\'ersic profile index ($n$) histograms for galaxies
   within FVS (solid lines) and outside FVS (dashed lines).  The same
   is shown in (b) for the sample of the brightest group galaxies.
        Dotted line in both panels indicates the threshold $n=3.5$.}
(c) Luminosity gap ($m_1-m_2$) histograms for groups with a late--type BGG 
   within FVS (solid lines) and outside FVS (dashed lines).  The same
   is shown in (d) for the groups with an early--type BGG.
   \label{F_01}
\end{figure}

In Figure \ref{F_01}(a), we show the S\'ersic profile index $n$
distributions for the galaxy samples inside (solid line) and outside
(dashed line) FVS.
We can notice that $n$ presents a similar behavior in all galaxies,
independently of the environment they are embedded in.
However, as previously mentioned, this parameter allows us to make a
morphological distinction between early and late--type galaxies
according to their surface brightness profiles.
In Figure \ref{F_01}(b) we show the S\'ersic index distribution
for the BGGs, inside (solid line) and outside (dashed line) FVS.
It can be seen that the distributions do not depart significantly from
each other, neither in the full sample nor the subsample of BGGs.
According to these distributions, and for the sake of obtaining two
samples of roughly the same size, we separate both galaxy samples by a
threshold of $n=3.5$, indicated by the dotted vertical line in panels (a) and (b).
This is not the conventional value adopted to discriminate
morphological characteristics of galaxies. 
Nevertheless, as our purpose is to study the brightest galaxy of each
group, which tends to be red, early and elliptical, we employ a less
restrictive value.
This allows us to distinguish between a sample dominated by groups
with an early--type brightest galaxy ($n>3.5$), and another sample
dominated by groups with a late--type brightest galaxy ($n<3.5$).
In an attempt to verify if this is a suitable classification, we
visually inspect the brightest group galaxies of each sample,
confirming that most of the brightest galaxies with $n>3.5$ can be
classified as early--type galaxies whereas those with $n<3.5$ are
likely to be late--type galaxies. 
Using this criteria, we obtain four samples of galaxies in groups
described in Table \ref{T_01}.
Sample $E_{in}$ is composed by galaxies in groups dominated by an
early--type galaxy and whithin a FVS; and sample $E_{out}$ is composed
by galaxies in groups dominated by an early--type galaxy not within a
FVS enviroment.
On the other hand, sample $L_{in}$ comprises the galaxies in groups
dominated by a late--type galaxy which also belong to a FVS and sample
$L_{out}$ contains the galaxies in groups outside FVS, dominated by a
late--type galaxy.

\begin{table} \begin{center}
\begin{tabular}{ cccc} 
 \hline
 within FVS  &  $E_{in}$ & $L_{in}$ & $Total_{in}$  \\
 $N_{grps}$ &   92 & 31 & 123 \\
 $N_{glxs}$  &  457 & 404 & 861 \\
 \hline
 outside FVS & $E_{out}$ & $L_{out}$ & $Total_{out}$  \\
 $N_{grps}$ &   270 & 102 & 372 \\
 $N_{glxs}$  &  1382 & 1238 & 2620 \\
 \hline
\end{tabular}%
\caption{Size of galaxy samples in groups dominated by an early--type
BGG $(E_{in/out})$ or a late--type BGG $(L_{in/out})$ inside and
outside FVS, respectively. }
\label{T_01} 
\end{center}
\end{table}
%
We have also computed the luminosity gap mentioned in Section \ref{lgap},
ie. the luminosity difference between the first and second ranked galaxies in the groups,
taking into account their morphologies and pertenence to FVS.
In the lower panels of Figure \ref{F_01}, we show separately the luminosity gap for groups with a late--type BGG (c)
and for groups with an early--type BGG (d), in FVS and elsewhere.
The median of the luminosity gap for groups dominated by late--type galaxies is approximately -0.36,
and is remarkably similar for groups in FVS and elsewhere.
For early types this luminosity gap median is -0.40 for groups
 in FVS and -0.30 elsewhere, a small difference at the 2 $\sigma$ level.
Taking into account these results, the groups analysed can not be considered as extremely dominant galaxies with satellites.
%

\subsection{Luminosity and multiplicity of group samples}

\begin{figure} 
   \centering
   \includegraphics[width=0.45\textwidth]{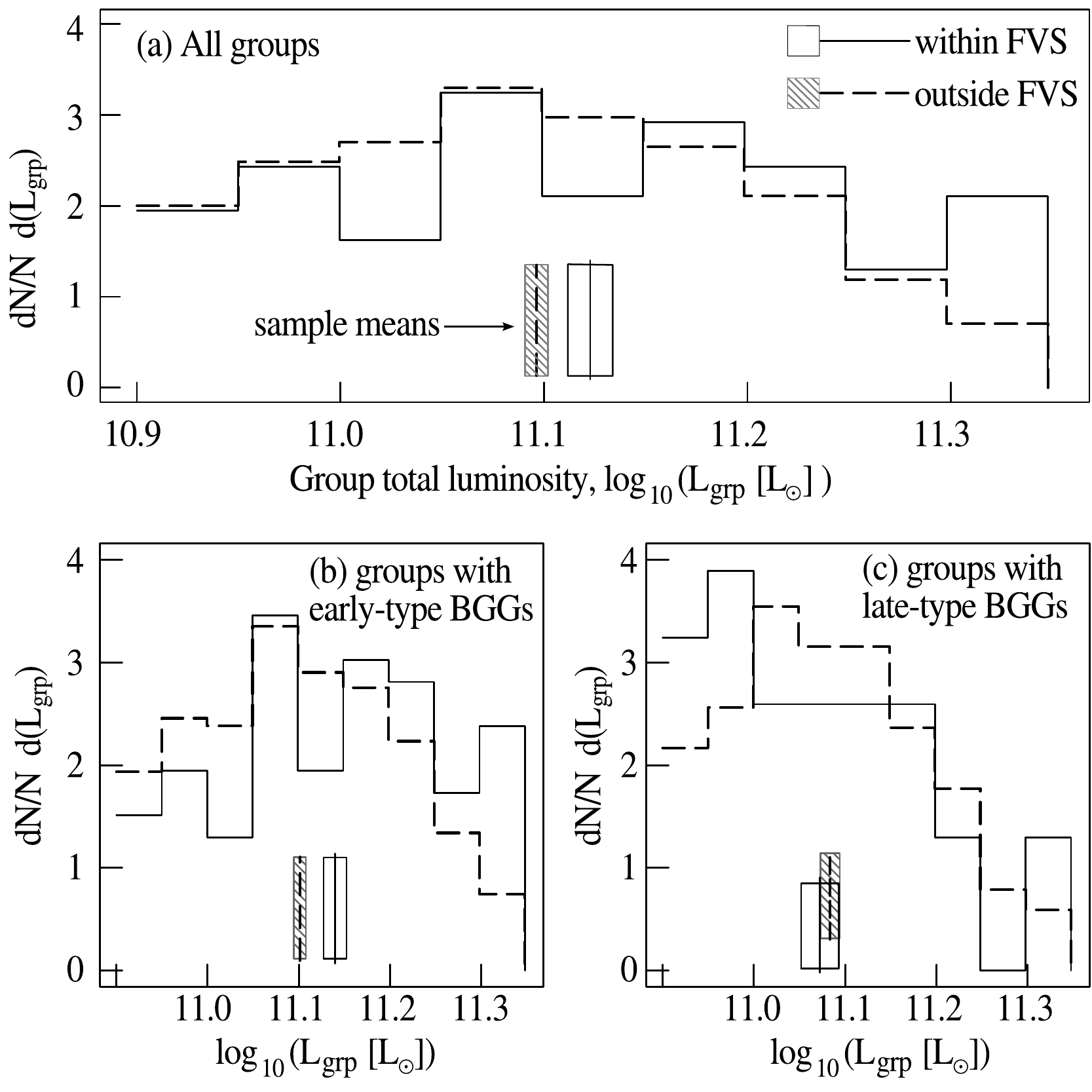}
   \caption{\textit{$M_r$}-band total luminosity histograms of groups within FVS
      (solid lines) and outside FVS (dashed lines), for (a) the total
      sample of groups, (b) groups with an early--type brightest galaxy
      ($E_{in/out}$ samples), and (c) groups with a late--type
      brightest galaxy ($L_{in/out}$ samples).  The means and mean
      uncertainties of the distributions are indicated by white and
      shaded boxes, for the samples of groups within and outside FVS,
   respectively.}
   \label{F_02}
\end{figure}

\begin{figure} 
   \centering
   \includegraphics[width=0.45\textwidth]{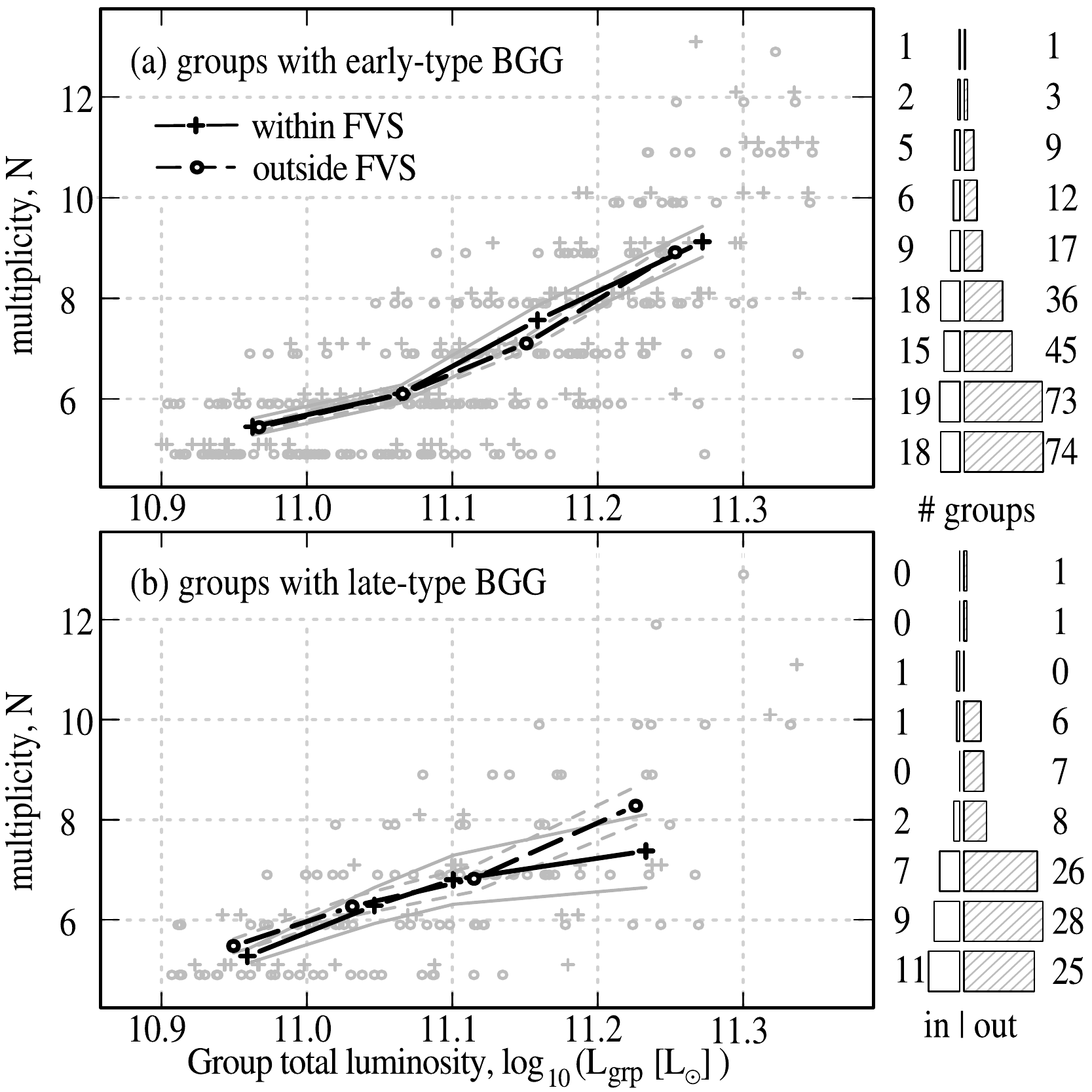}
   \caption{Mean multiplicity of groups within (solid lines) and
      outside (dashed lines) FVS, as a function of total group
      \textit{$M_r$}-band luminosity, separately for (a, upper panel)
      groups with an early--type brightest galaxy, and (b, botom panel) 
      groups with a late--type brightest galaxy.
      The individual values of total group luminosity and multiplicity
      are represented for each group with grey cross symbols and circles, 
      for groups within and outside FVS, respectively.
      Bar charts with the number of groups within (left bars) and 
      outside (right, shaded bars) FVS per multiplicity value are shown 
      on the right.
   }
   \label{F_03}
\end{figure}

\begin{figure} 
   \centering
   \includegraphics[width=0.45\textwidth]{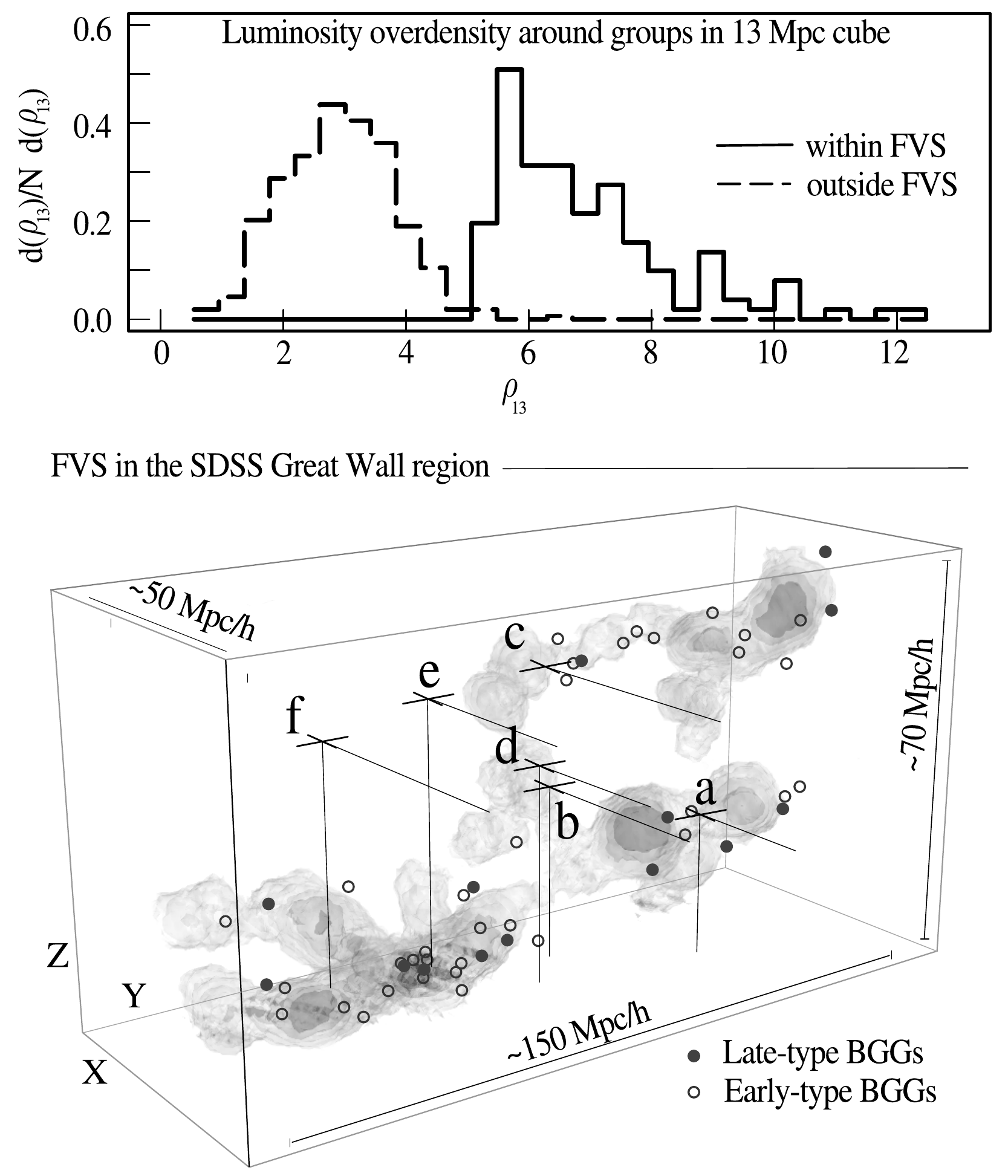}
   \caption{Top panel: distribution of local densities in 13 Mpc cubic cells centered in each group of the total
sample. Solid line corresponds to groups in FVS and dashed line to groups elsewhere.
Bottom panel: luminosity density of the FVS associated to the Sloan Great Wall. The positions are shown in a cartesian
system associated to the equatorial coordinate system. Lighter shades of grey correspond to higher densities, and white corresponds to regions outside the Great Wall.
The member groups are superimposed showing separately those dominated by early and late type galaxies.
The crosses mark the centre of mass of the superclusters identified by \citet{Liivamagi:2010} in the same region:
(a) SCL 184+003+0077; (b) SCL 202-001+0084; (c) SCL 187+008+0089; (d) SCL 189+003+0086; (e) SCL 196+011+0086; (f) SCL 198+007+0093.
  }
   \label{GG}
\end{figure}

Among group properties, the total group luminosity is one of the most
relevant quantities, since it strongly correlates with group total
mass and therefore with galaxy member properties.  
\citet{Lacerna:2014} study the properties of halo-central galaxies and
compare central galaxies in groups (host halos containing satellites)
to isolated central galaxies (host halos without satellites).
They find that group central galaxies are redder and less star forming
than field central galaxies, although the color and specific star
formation rate distributions at the same stellar mass are comparable.
The authors argue that central galaxies which assembled in dense
environments like groups or clusters, tend to have larger masses.
Figure \ref{F_02} shows the luminosity distribution for groups
inside (solid line) and outside (dashed line) FVS.
Vertical lines (solid line for galaxies inside FVS and dashed line for
galaxies elsewhere) indicate the sample means, and the boxes represent
the error estimated by standard deviation.
The total group luminosity distribution is shown in the upper panel,
and those corresponding to samples $L_{in/out}$ and $E_{in/out}$ in
the lower panels, exhibiting that groups in FVS are more luminous than
groups elsewhere.
This enforces the fact that, in order to compare galaxies within and
outside FVS, we have to take into account whether the properties of
the group samples in FVS and elsewere are similar. 
Firstly, we have checked that the distance distribution of group whithin
FVS and elsewhere are similar, assuring no redshift dependent systematic 
effects.
Besides, groups in samples $L_{in}$ and $L_{out}$ do not show significant
luminosity differences respect to those between $E_{in}$ and $E_{out}$ samples. 
Moreover, the multiplicity could have a significant effect on the
comparison of galaxy groups.
In the Figure \ref{F_03} we show the mean group multiplicity $N$ per bin
of group luminosity, inside and outside FVS, comparing samples
$E_{in}$ (solid line) with $E_{out}$ (dashed line) in the upper panel
and $L_{in}$ (solid line) with $L_{out}$ (dashed line) in the lower
panel.
Bins of group luminosity are taken on an equal-number basis.
As can be seen, in both cases the multiplicity has a similar behavior
therefore we do not expect any systematic effect on the group
comparison, except maybe on the highest luminosity groups dominated by
late--type BGGs.
\\ \\
In figure \ref{GG} we show the luminosity density of the FVS associated to the Sloan Great Wall.
The member groups are superimposed showing separately those dominated by early and late--type galaxies.
The plot shows a similar distribution although with a tendency of groups dominated by late--type galaxies to lie in the outskirts.
In the top of the figure it is shown the distribution of local densities in 13 Mpc cubic cells centered in each group of the total
sample, where it can be seen that high density groups correspond approximately to those in FVS although there is a small fraction
of groups with local high densities that are not member of FVS.
With the use of FVS in our study, we are able to distinguish these regions of high global densities but that do not belong to superstructures.

%
%
%
%
%
\section{Properties of Group Galaxies} \label{S_properties}
%
In order to study the possible environmental effects on galaxies in
groups, we analyse variations of properties associated to the star--formation
activity, depending on whether they reside or not in FVS.
For this aim, we have used the $u-r$ color index that provides a suitable
indication of recent episodes of star formation.
In spite of their larger uncertainties, we have used the $u-r$ color index
instead of $g-r$ since we have shown in \citet{Lambas:2012} that star formation induction
can be succesfuly detected using this index.

In addition, we have also analysed the stellar mass content and star formation rate provided in 
the SDSS database.
The concentration parameter $R_{50}/R_{90}$ also provides a suitable measure of the light concentration.

\subsection{Properties according to the galaxy ranking}
\label{g_ranking}

\begin{figure} 
   \centering
   \includegraphics[width=0.45\textwidth]{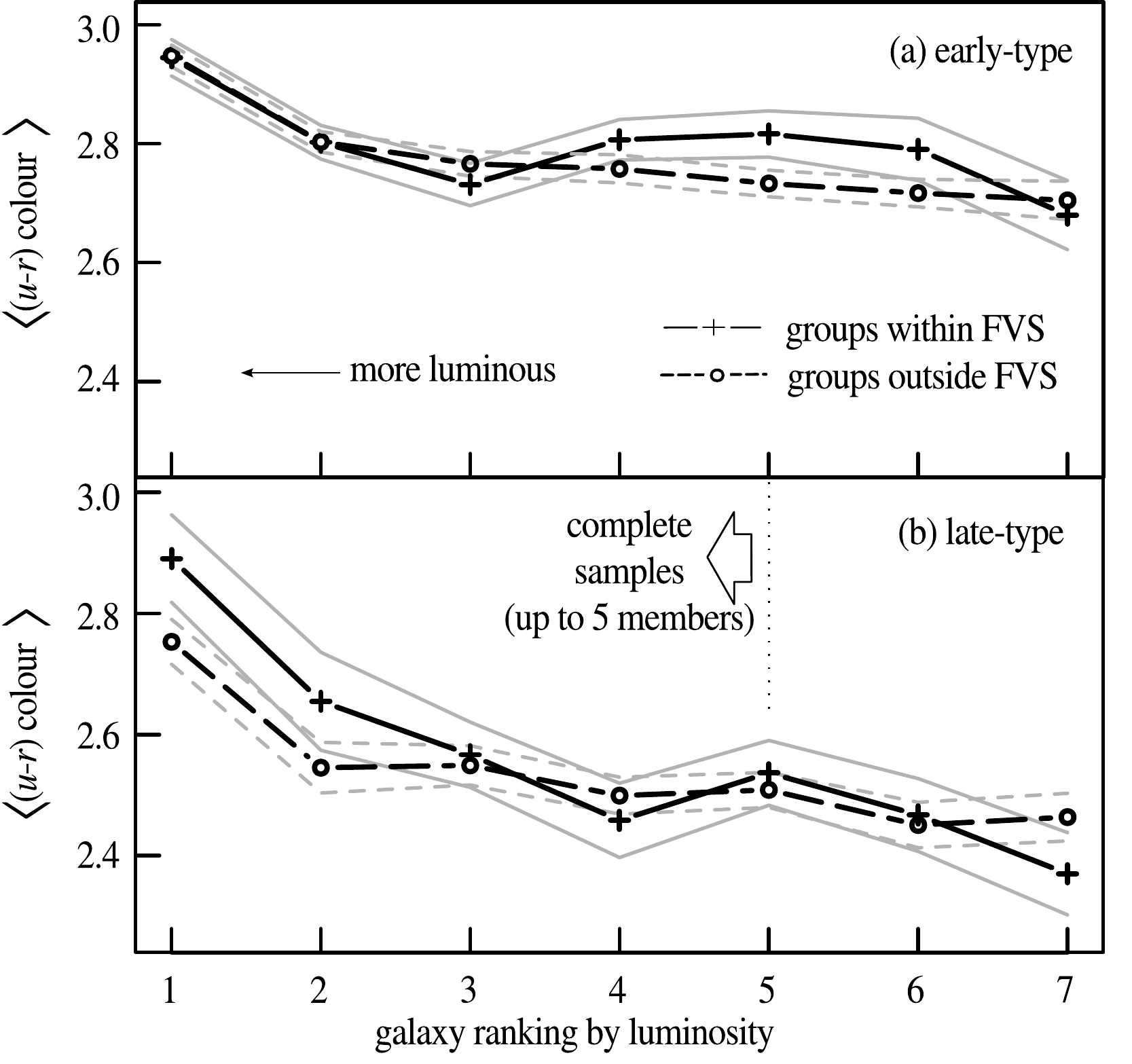}
   \caption{Mean \textit{u-r} colors of (a) early--type and (b)
      late--type galaxies as a function of galaxy ranking, for groups
      within FVS (solid lines, cross symbols) and outside FVS (dashed
      lines, circles).  Grey lines represent 1-$\sigma$
   uncertainties.} 
   \label{F_04}
\end{figure}

\begin{figure}
   \centering
   \includegraphics[width=0.45\textwidth]{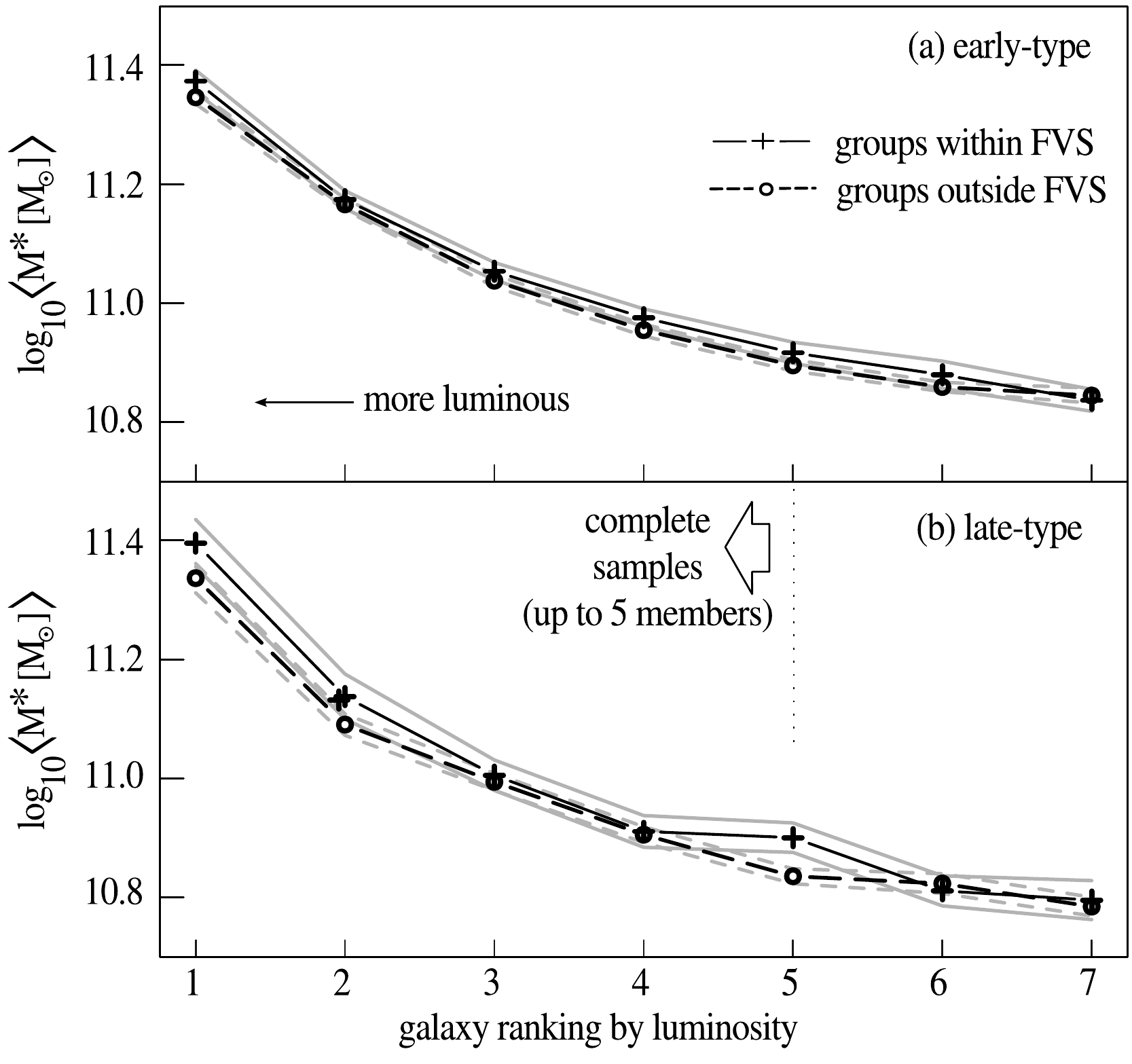}
   \caption{Mean stellar mass $M^{*}$ of (a) early--type and (b)
      late--type galaxies as a function of galaxy ranking, for groups
      within FVS (solid lines, cross symbols) and outside FVS (dashed
      lines, circles).  Grey lines represent 1-$\sigma$ uncertainties.
   } 
   \label{F_05}
\end{figure}

%
We have studied the behaviour of mean \textit{u-r} colors and
stellar masses of the different samples as a function of galaxy
ranking.
In Figure \ref{F_04} we show the mean \textit{u-r} colors of samples
$E_{in}$ (solid line) and $E_{out}$ (dashed line) in the top panel,
and of samples $L_{in}$ (solid line) and $L_{out}$ (dashed line) in
the bottom panel.
All groups contribute with one galaxy up to ranking 5, but for higher
ranking the averaged values are computed using only the richer groups.
The vertical dashed lines in the Figure \ref{F_04} indicate the
completness of the group samples.
The environmental influence over galaxies in groups dominated by a
late--type BGG is noticeable on the two more luminous galaxies (lower
panel of Figure \ref{F_04}), which are redder if they belong to a
group within FVS, with 1-$\sigma$ significance.
Conversely, this effect is not noticeable for galaxies in groups
dominated by an early--type BGG.  
In the top panel of Figure \ref{F_05} we show the variation of the
mean stellar mass of samples $E_{in}$ (solid line) and $E_{out}$
(dashed line) as a function of galaxy ranking.
Similary, the results for samples $L_{in}$ (solid line) and $L_{out}$
(dashed line) are shown in the bottom panel of this Figure.
In both panels it can be seen a similar trend, however there is a
slight difference between samples $L_{in}$ and $L_{out}$.  
This suggests that the environmental influence on stellar mass is just
detectable for galaxies in groups dominated by a late--type BGG. 
By inspection of both figures, it can be seen that a significant
effect appears in the first--ranked galaxy and decays for higher
ranked galaxies.
%
%
As analysed in subsection \ref{lgap}, it is interesting to notice that even 
though the luminosity gap in groups is not large, 
it is still the central galaxy the one that shows differences in its properties.

\normalfont

According to these results, hereafter we will explore in detail the
effect of the FVS on the galaxy properties considering only the
first--ranked galaxies (BGGs).
%
%
%
\subsection{Brightest Group Galaxy and large--scale environment} \label{first_ranked}
%
To ensure that the features pointed out in the previous section are
due to the influence of the the large--scale environment defined by
FVS, we restrict the following analysis to bins with the same total
group luminosity inside and outside FVS, characterizing the local
environment of main galaxies by the luminosity of their host groups.
According to the results of Section \ref{g_ranking}, galaxies other
than the first--ranked do not show a large--scale environment
dependence, thus in this section we study only the properties of the
BGGs.

\begin{figure*} 
   \centering
   \includegraphics[width=\textwidth]{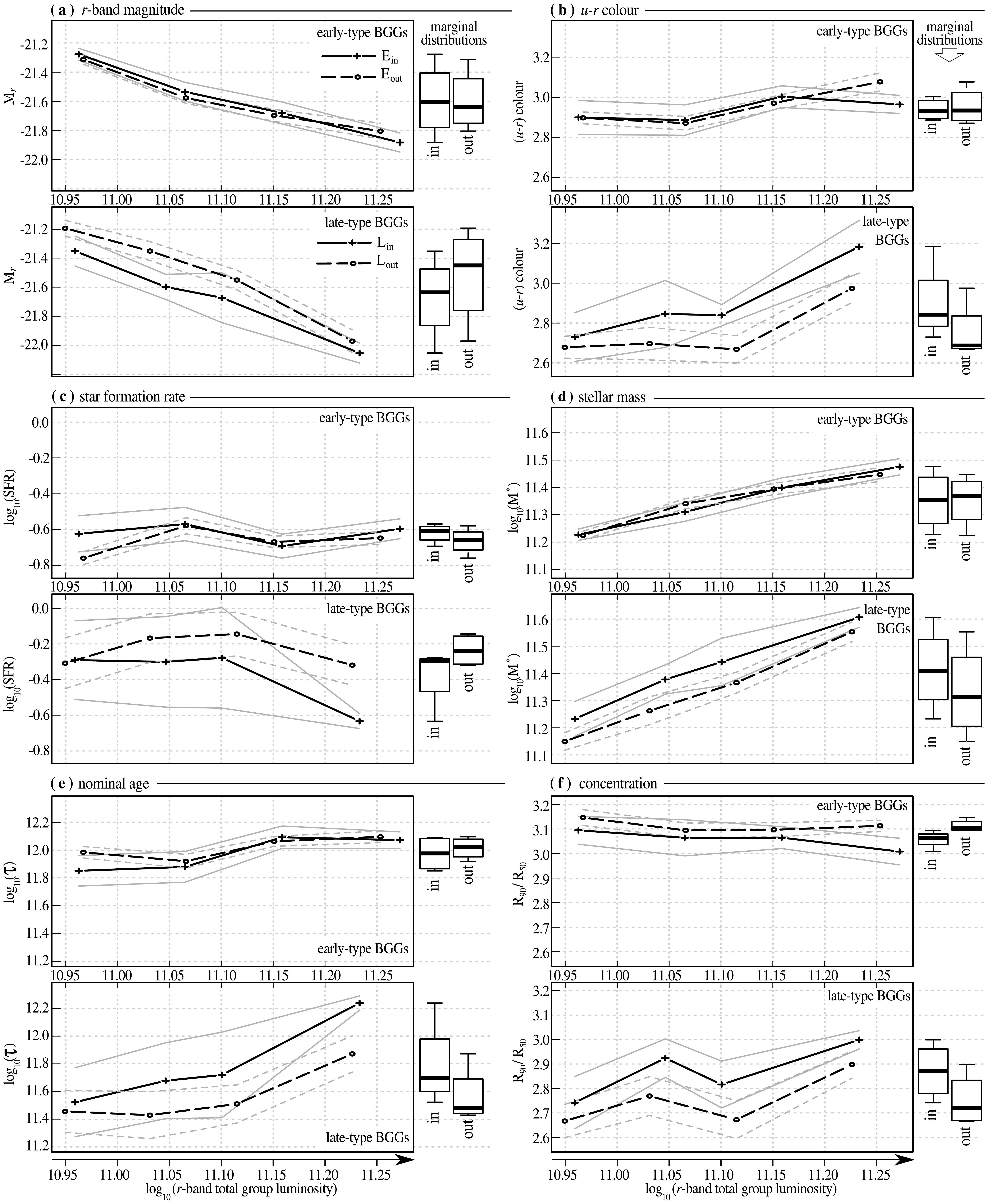}
   \caption{Averaged properties of early--type (upper subpanels) and
      late--type (lower subpanels) BGGs in FVS (solid lines) and
      elsewhere (dashed lines), as a function of total group
      luminosity.  1-$\sigma$ uncertainties are shown with solid 
      and dashed gray lines, respectively.  The properties shown in
      this Figure are a) r-band magnitude,
      b) \textit{u-r} colour,
      c) star formation rate,
      d) stellar mass,
      e) nominal age
      and f) concentation index.
      In all panels L$_{in}$/L$_{out}$ and E$_{in}$/E$_{out}$ refer 
      to late--type BGGs within/outside FVS
      and early--type BGGs within/outside FVS, respectively.
      Box plots on the right hand side of each panel
      show the quartiles and extrema of the correponding marginal distributions.}
   \label{F_06}
\end{figure*}

Besides the comparison on equal total luminosity groups, we aim at
studying BGGs with similar morphology.
Therefore, we analyse averaged properties of the BGGs as a function of
the \mbox{\textit{r}--band} total group luminosity, as shown in the
different panels of Figure \ref{F_06}.
In this Figure we show the \textit{r}-band magnitude, \textit{u-r}
colour, star formation rate, stellar mass, nominal time--scale (the
ratio of stellar mass to star formation rate) and concentation index,
in the panels (a) to (f), respectively.
In all cases, samples $E_{in}$ and $L_{in}$ are shown with solid
lines, and samples $E_{out}$ and $L_{out}$ with dashed lines.  Grey
lines represent 1-$\sigma$ uncertainties of the averaged properties in
each luminosity bin.
In general, the difference between groups in high density peaks and
groups not belonging to FVS is noticeable with a 1-$\sigma$
significance level, for the samples of groups with a late--type BGG
($L_{in}$ and $L_{out}$), over the entire group luminosity range
(lower panel of Figure \ref{F_06}).
This difference, however, is not significant for groups with an early--type
BGG, as can be seen in the upper subpanels of the Figure
\ref{F_06}.

Regarding to the \textit{$M_r$} dependence on luminosity, it is clearly seen
that the late--type BGGs in FVS are more luminous than their
counterparts outside FVS (see Figure \ref{F_06}(a)).
This trend is also evident from the marginal distributions (box plots
at the right), where the medians of the early--type BGGs samples,
$E_{in}$ and $E_{out}$, are very similar, while late--type BGGs are
noticeably more luminous when they are within FVS.
The same trend can be noticed in the mean \textit{u-r} colors, in
Figure \ref{F_06}(b), where it can be seen that the late--type BGGs in
FVS are redder than those outside FVS.
Meanwhile, the early--type BGGs behave similarly regardless of their
environment.
With respect to the mean star formation rate (SFR) shown in Figure
\ref{F_06}(c), there is also a tendency of the late--type BGGs to be
less star--forming in FVS.
Despite the more extended error bands, the difference in the mean SFR
holds over the entire luminosity range (except may be for the less
luminous groups) for late--type BGGs, but it is close to zero for the
early--type BGG samples.
For the less luminous groups, this effect inverts, showing a
difference for the early--type BGGs, but not for the late--type.
Figure \ref{F_06}(d) also shows that the mean stellar mass $M^*$ for
late--type BGGs is greater for galaxies in FVS than elsewhere.
This difference also holds for groups in the whole luminosity range,
and once again, is not significant for the early--type BGGs residing
in groups of roughly the same total luminosity.

In order to assign an indicator of the galaxy time--scale we define
the parameter $\tau = M^*/SFR$, which provides an estimate of the
time--scale for the formation of the total stellar mass at the present
rate of star formation.
This parameter is displayed in Figure \ref{F_06}(e) where, in spite of
the errors, the late--type BGGs exhibit larger time--scales when they
are in FVS.
On the other hand, the concentration index correlates with galaxy
morphological type \citep{Shimasaku:2001, Strateva:2001,
Nakamura:2003} and is defined as the ratio of the two Petrosian radii
$C = R_{90}/R_{50}$ measured in the \textit{r}--band, where $R_{90}$ and
$R_{50}$ are radii corresponding to the apertures which include 90 per
cent and 50 per cent of the Petrosian flux, respectively.
In Figure \ref{F_06}(f) it can be seen that the early--type BGGs in FVS
are slightly less concentrated than their counterparts outside FVS.
For the late--type BGGs this effect is more remarkable and opposed;
early--type galaxies are less concentrated if they lay within FVS,
late--type BGGs are more concentrated when they belong to FVS.
Finally, we have also explored samples of BGGs with equal luminosity
distribution in FVS and elsewhere to check if the trends in color 
index, star--formation, age and concentration remain, or are due to
the larger luminosities of BGGs in FVS. 
We find that the previously observed differences are similarly
detected for these equal luminosity subsamples, reinforcing the astrophysical
signals of the large--scale environment effect.

We acknowledge that contamination by background/foreground galaxies is likely to affect the total luminosities of
a fraction of less than about 8 per cent of the groups \citep{Zapata:2009}.
However, we expect this contamination to be small given the relatively large number of true member galaxies and that this
effect is expected to be uncorrelated with group properties. 
Besides, we argue that the previous results are not likely to be affected 
since our analysis relies on the brightest group members for which we expect a 
negligible contamination.

%

\section{Quantifying the large--scale environment dependence of early and late--type
Brightest Group Galaxies} \label{S_BGGtype}

In order to quantify the large--scale environment dependence of early
and late--type BGGs properties regardless the host group luminosity,
we have considered departures from the mean values of such properties
at a given group luminosity interval.
For the complete sample of early (late) type BGGs, regardless of their
location on the large--scale structure, we perform a linear fit of
each property as a function of group luminosity.
We then estimate the residuals of the fit, separately for the galaxies
inside and outside FVS.
If the FVS environment has no influence over galaxy properties, then
these residuals should be distributed similarly.
Otherwise, any trend on the mean of the residual values inside and
outside FVS can be considered as an indicator of large--scale effects.
This procedure resembles that of the principal component analysis,
except that residuals in the axis measuring the galaxy property are
used instead of the second principal component \citep{Jolliffe:2005}.

\begin{figure}
   \centering
   \includegraphics[width=\columnwidth]{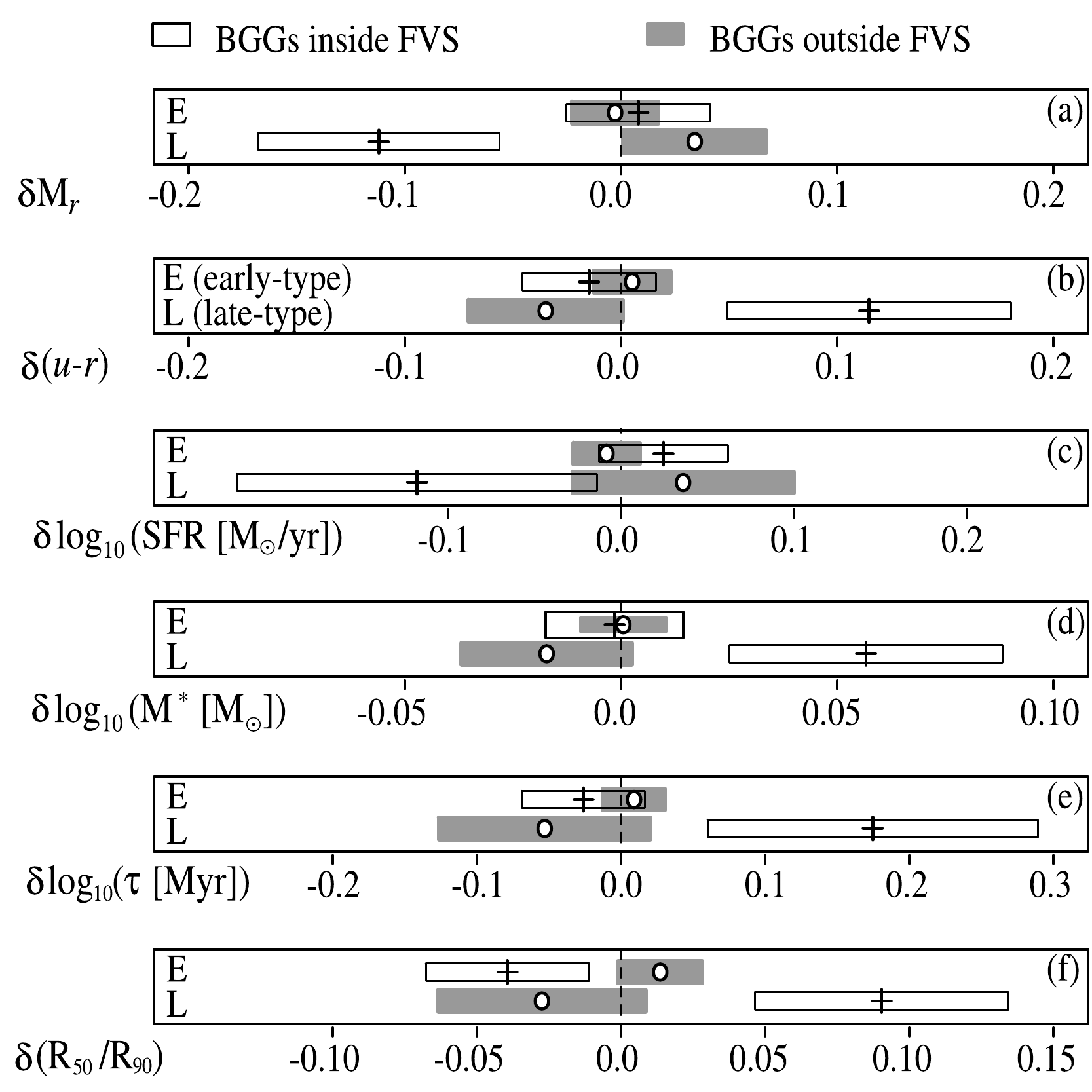}
   \caption{Mean residuals and their standard errors for early--type
      (E) and late--type (L) BGGs.  The residuals are computed for (a)
      \mbox{\textit{$M_r$}-band} luminosity, (b) \textit{u-r} colour
      index, (c) star formation rate, (d) stellar mass, (e) nominal
      galaxy time--scale, and (f) concentration index.
      White boxes and cross symbols correspond to galaxies in groups
   within FVS ($E_{in}$ and $L_{in}$ samples), grey boxes and circles
correspond to galaxies in groups outside FVS ($E_{out}$ and $L_{out}$
samples).}            
   \label{F_07}
\end{figure}

The mean residuals of the \textit{r}-band absolute magnitude, the
\textit{u-r} color, the concentration index $R_{50}/R_{90}$, the
logarithm of the star formation rate (\mbox{log$_{10}$(SFR)}), the
stellar mass (\mbox{log$_{10}$(stellar mass)}), and galaxy SFR
time--scale \mbox{(log$_{10}$($\tau$))} are shown in Figure
\ref{F_07}.  
Cross symbols and white boxes indicate BGGs inside FVS, and circles
and grey boxes are used for BGGs outside FVS.  
The width of boxed regions around each symbol indicates the standard
deviation.
It can be seen that in most cases late--type BGGs exhibit more
significant differences according to their the large--scale
environment while early--type BGGs are likely to belong to a common
population, irrespective of whether they belong or not to a FVS.
A remarkable exception to this uniformity in the early--type behaviour
is given by the concentration parameter, we recall this later below.
First, lets consider the panel (a) of Figure \ref{F_07}. 
As it can be seen, early--type galaxies do not exhibit significant
deferences in the \textit{r}-band magnitudes. 
For the late--type galaxies there are statistically significant
differences indicating that BGGs outside FVS tend to be brighter than
their corresponding sample in FVS. 
Secondly, in panel (b) of this figure, we find that late--type
galaxies show noticeable differences in color residuals, depending on
their FVS membership.  
Moreover late--type BGGs outside FVS have bluer residuals in
comparison with systems inside FVS. 
However, it should be noticed that this difference is dominated by
BGGs in FVS, since the mean of the residuals for this sample is $0.11
\pm 0.06$ redder than the total mean, whereas the mean colour of BGGs
outside FVS is not statitically different from the mean colour of the
total sample.  
In contrast for early--type systems no dependency can be observed.

Following the analysis of Figure \ref{F_07}, in panel (c) we observe a
marginally enhancement in the star formation rate for late--type
systems outside FVS.  
We agree however that the obtained values lie inside the one standard
deviation level significance. 
Even smaller differences are shown by early--type BGGs samples
regardless of their FVS membership.
Panel (d) shows a notable increment in stellar mass residuals for
late--type BGGs inside superstructures. 
It should be noticed that residuals are computed over the logarithm of
the mass, indicating that roughly a $10\%$ of excess in stellar mass
residuals for systems belonging to FVS respecting to galaxies outside
superstructures. 
This signal goes in the same direction than the observed in panel (e)
of Figure \ref{F_07}. 
In this panel we present the results for the logarithm of the star
formation time--scale, depending on BGGs morphology and FVS
membership. 
Here again we see an statistically significant difference for
late--type systems whether they lay or not in FVS. 
Brightest late--type galaxies inside superstructures seem to be $30\%$
older than its corresponding sample outside FVS.
Finally, in panel (f) we show the results for residuals on the
concentration parameter ($R_{50}/R_{90}$). 
Remarkably, for this property we observe environment dependences for
both galaxy morphologies. 
Early--type galaxies inside FVS exhibit notably less concentrated
residuals than their analogous sample of field galaxies. 
This suggests that these systems may be have been engaged in more
frequently galaxy mergers than the field early--type galaxies. 
In contrast for late--type BGGs, we observe positive excess in
concentration residuals in comparison to FVS member galaxies.

To quantify the statistical significance of these results, we consider
a random set of subsamples of groups not belonging to FVS dominated by
late--type galaxies and we compute the number of occurrences where the
mean of the residuals of the different brightest galaxy properties is
greater or equal than that corresponding to the dominant galaxy in FVS
groups.
We apply a resampling--method estimation using $10^4$ random
realizations of subsamples of groups not belonging to FVS finding that
the difference between the sample averages of inner and outer
late--type galaxies is statistically significant.
The null hypothesis of the samples of late--type galaxies in and out
FVS being part of the same parent distribution can be disproved up to
a significance level of 90 per cent.

We confirm that late--type BGGs are brighter, redder, less
star--forming, with a higher content of stellar mass, a larger
time--scale for star formation, and more concentrated when they are
inhabiting FVS than elsewhere.
It is possible that these differences are due to the different
large--scale amplitude of clustering of equal luminosity groups inside
and outside FVS.
However, we have tested different virial mass thresholds for group
samples outside FVS to analyse whether their BGGs have properties
comparable to the total group sample in FVS.
We find that outside FVS, groups with masses as large as
\mbox{$M_{vir} > 10 ^{14.4} M_{\odot}$} have to be considered so that
their BGG properties are similar to the group sample in FVS. 
The relevance of the results shown in this section can be assesed by
the fact that this high mass group subsample outside FVS roughly
corresponds to the 10 per cent of more massive groups so that the
large--scale environment does play a fundamental role in setting the
properties of BGGs.
%
%
\section{Discussion and Conclusions}\label{S_conclusus}
%
As described in Section \ref{S_intro}, there are several previous
studies focused on the effects of the large--scale structure on galaxy
properties \citep[e.g.,][and references therein]{Binggeli:1982,
   Einasto:2003b,Einasto:2005, Donoso:2006, Einasto:2007c, Crain:2009,
White:2010, Croft:2012, Yaryura:2012, Einasto:2014}.
Within this framework, our main concern is to investigate and quantify
the large--scale dependence of the properties of galaxies in groups.
To this end, we have analysed properties of galaxies that reside in
similar local environments, by comparing samples in equal group total
luminosity (mass) intervals.
Our procedure has the advantage of separating the local and the global
effects on group members.

The properties of the brightest group galaxy (BGG) and their host
groups are strongly correlated
\citep{Agustsson:2010,Guo:2011,Lares:2011,Wang:2012,Wang:2014}.
We can consider that early--type BGGs reside typically in more
dynamically relaxed groups than those where the BGG is of late--type.
Taking this into account, we characterized the BGGs according to their
S\'ersic index $n$ (used as an estimator of morphology, see section
\ref{galaxy_samples} for details), and we defined four subsamples of
group galaxies: $E_{in-out}$ dominated by an early--type BGG, and
$L_{in-out}$, dominated by a late--type BGG, residing in
superstructures or elsewhere, respectively.
Besides, we have ordered the member galaxies according to their
luminosity in order to asses the importance of the ranking within the
groups in the response to the large--scale associated effects.

We can derive several conclusions from our study,
concerning the influence of the structure at large scales on the
properties of individual galaxies within groups, independently of the
local environment given by the group properties.
The main result is that the brightest galaxy within groups is by far more
affected than the rest of group members.
In fact, the change in the properties of galaxies beyond rank three in
luminosity are negligible, although they exhibit the influence of the
local environment given by the mass of the group they inhabit.
The effects on the second brightest galaxy are marginally detected.
The influence of the large scale on the BGGs does manifest on their
colours, star formation rate, stellar mass content, concentration
index and nominal age, and strongly depend on the galaxy type.
While the properties of late--type BGGs differ significantly
according to the
large--scale structure, the averaged values of the properties of
early--type galaxies are consistent within uncertainties.
Late--type
brightest group galaxies show
statistically significant higher luminosities and stellar masses,
redder \textit{u-r} colours, lower star formation activity and larger
star--formation time--scale when embedded in superstructures, and
regardless of the group local environment. 
Our analysis comprises tests against the dependence on the host group
luminosity, so that the effects of local environment, given by the
properties of the group, and large--scale environment, given by the
properties of the FVS, are disentangled.
We argue that group brightest member properties are not
only determined by the host halo, but also by the large--scale
structure.
Since the differences are significant only for late--type galaxies, 
they could be produced by the dependence of the accretion process onto
the brightest galaxies, which in turn is conditioned by the mass
overdensity at early times, and thus manifested today at large scales.

Previous works by different authors have analysed the correlation between group environment and the
properties of their member galaxies.
\normalfont
\citet{Lietzen:2012} conclude that member galaxies in groups with
similar richness are more likely to be passive if they are in
superclusters.  \citet{Einasto:2014} find that supercluster morphology
affects the galaxy population: filament--type superclusters contain a
larger fraction of red, early and low star--forming galaxies than
spider--type superclusters.
Also, they find that blue, high SFR galaxies have lower environmental
densities (defined within an $8 Mpc/h$ smoothing length) than red, low
SFR galaxies in both types of superclusters.  
\citet{Einasto:2011} study the role of the first ranked galaxies in groups
 in the region of the Sloan Great Wall finding that the spatial distribution
 of groups in the context of their hosting superclusters depend on whether
 they have an elliptical or spiral brightest galaxy. \normalfont
\citet{Luparello:2013} compare equal global luminosity groups (a proxy
of the total mass) concluding that groups in superstructures formed
earlier than groups located in lower density regions.
Consistently, we conclude that late--type BGGs inhabiting
superstructures show 10 per cent higher luminosities, 10 per cent
stellar mass excess, 0.11 redder \textit{u-r} colours, significantly lower star formation
activity and with a 30 per cent longer star--formation time--scales with respect to the means
of the total samples (see section \ref{S_BGGtype} for details of these calculations).
These differences are negligible when considering lower luminosity
galaxies, indicating that the effects are related to the particular
role of the brightest group galaxy.

We argue that these differences 
are only found in late--type BGGs
because of their formation history.
The observed signals support a scenario where the gas accretion
via mergers
(relative to that of the stars) into the BGG is more important when
the groups are located outside FVS.
The accretion occuring in groups inhabiting FVS could be
dominated by local dynamical processes such as galaxy harassment and
gas extrangulation, resulting in a negligible effect of the FVS
environment over low ranked galaxies.
The more efficient accretion of gas onto the BGG not residing in FVS
would be associated to the effects on late--type BGGs. Although this
effect could also be present on early--type BGGs, their older ages
could have erased the effect on these objects.
However, the concentration parameter of early--type BGGs in FVS is
significantly lower suggesting more recent merger events.
In sum, our results indicate a significantly larger fraction of dry
mergers compared to wet mergers, occuring onto BGGs in FVS.
%
%
\section*{acknowledgements}
We thank the Referee, Maret Einasto, for her through review and
highly appreciate the comments and suggestions, which greatly improved
this work.
This work was partially supported by the
Consejo Nacional de Investigaciones Cient\'{\i}ficas y T\'ecnicas
(CONICET), and the Secretar\'{\i}a de Ciencia y Tecnolog\'{\i}a,
Universidad Nacional de C\'ordoba, Argentina.
NP was supported by Fondecyt regular 1110328 and BASAL CATA PFB-06.
Funding for the SDSS and SDSS--II has been provided by the Alfred P.
Sloan Foundation, the Participating Institutions, the National Science
Foundation, the U.S. Department of Energy, the National Aeronautics
and Space Administration, the Japanese Monbukagakusho, the Max Planck
Society, and the Higher Education Funding Council for England. The
SDSS Web Site is http://www.sdss.org/.
The SDSS is managed by the Astrophysical Research Consortium for the
Participating Institutions. The of the Royal Astronomical Society
Participating Institutions are the American Museum of Natural History,
Astrophysical Institute Potsdam, University of Basel, University of
Cambridge, Case Western Reserve University, University of Chicago,
Drexel University, Fermilab, the Institute for Advanced Study, the
Japan Participation Group, Johns Hopkins University, the Joint
Institute for Nuclear Astrophysics, the Kavli Institute for Particle
Astrophysics and Cosmology, the Korean Scientist Group, the Chinese
Academy of Sciences (LAMOST), Los Alamos National Laboratory, the
Max-Planck-Institute for Astronomy (MPIA), the Max-Planck-Institute
for Astrophysics (MPA), New Mexico State University, Ohio State
University, University of Pittsburgh, University of Portsmouth,
Princeton University, the United States Naval Observatory, and the
University of Washington.   
%
%
Plots are made using R software and post-processed with Inkscape.

\bibliographystyle{mn2e}
\bibliography{Bibliography.bib}

\end{document}